\documentclass[
aip,
 amsmath,amssymb,
 reprint,%
]{revtex4-1}
\usepackage{graphics}
\usepackage{graphicx}
\usepackage{epsfig}
\usepackage{amssymb}
\usepackage{amsthm}
\usepackage{xcolor}
\usepackage{lipsum}
\usepackage{amsmath}

\usepackage{upgreek}

\usepackage{graphicx}% Include figure files
\usepackage{dcolumn}% Align table columns on decimal point
\usepackage{bm}% bold math
\usepackage[utf8]{inputenc}
\usepackage[T1]{fontenc}
\usepackage{mathptmx}
\usepackage{etoolbox}

\makeatletter
\def\@email#1#2{%
 \endgroup
 \patchcmd{\titleblock@produce}
  {\frontmatter@RRAPformat}
  {\frontmatter@RRAPformat{\produce@RRAP{*#1\href{mailto:#2}{#2}}}\frontmatter@RRAPformat}
  {}{}
}%
\makeatother

\begin{document}

\title{Dimensional Crossover of Microscopic Magnetic Metasurfaces for Magnetic Field Amplification}

\author{N.~Lejeune}
\affiliation{Experimental Physics of Nanostructured Materials, Department of Physics, Universit\'{e} de Li\`{e}ge, B-4000 Sart Tilman, Belgium.\\}
 
\author{E.~Fourneau}
\affiliation{Experimental Physics of Nanostructured Materials, Department of Physics, Universit\'{e} de Li\`{e}ge, B-4000 Sart Tilman, Belgium.\\} 

\author{A.~Barrera}
\affiliation{Insititut de Ci\`encia de Materials de Barcelona, ICMAB-CSIC, Campus de la UAB 08193 Bellaterra, Spain.\\} 

\author{O.~Morris}
\affiliation{Centre for Nanoscience and Nanotechnology, Department of Physics, University of Bath, Bath, BA2 7AY, United Kingdom.\\}
 
\author{O.~Leonard}
\affiliation{Centre for Nanoscience and Nanotechnology, Department of Physics, University of Bath, Bath, BA2 7AY, United Kingdom.\\}  
  
\author{J.~A.~Arregi}
\affiliation{CEITEC BUT, Brno University of Technology, Purky\~{n}ova 123, 612 00 Brno, Czech Republic.\\}

% \author{M. Stano}
% \affiliation{Institute of Physical Engineering, Brno University of Technology, Technick\`{a} 2, 616 69 Brno, Czechia.\\}

\author{C.~Navau}
\affiliation{Grup d’Electromagnetisme, Departament de Fisica, Universitat Autonoma de Barcelona, 08193 Bellaterra, Barcelona, Spain.\\}

\author{V.~Uhl{\'{i}}ř}
\affiliation{CEITEC BUT, Brno University of Technology, Purky\~{n}ova 123, 612 00 Brno, Czech Republic.\\}
\affiliation{Institute of Physical Engineering, Brno University of Technology, Technická 2, 616 69 Brno, Czechia.\\}

\author{S.~Bending}
\affiliation{Centre for Nanoscience and Nanotechnology, Department of Physics, University of Bath, Bath, BA2 7AY, United Kingdom.\\}

\author{A.~Palau}
\email[Corresponding author: ]{palau@icmab.es}
\affiliation{Insititut de Ci\`encia de Materials de Barcelona, ICMAB-CSIC, Campus de la UAB 08193 Bellaterra, Spain.\\} 

\author{A.~V.~Silhanek}
\email[Corresponding author: ]{asilhanek@uliege.be}
\affiliation{Experimental Physics of Nanostructured Materials, Department of Physics, Universit\'{e} de Li\`{e}ge, B-4000 Sart Tilman, Belgium.\\} 

\date{\today} % Leave empty to omit a date

\begin{abstract}

Transformation optics applied to low frequency magnetic systems has been recently implemented to design magnetic field concentrators and cloaks with superior performance. Although this achievement has been amply demonstrated theoretically and experimentally in bulk 3D macrostructures, the performance of these devices at low dimensions remains an open question. In this work, we numerically investigate the non-monotonic evolution of the gain of a magnetic metamaterial field concentrator as the axial dimension is progressively shrunk. In particular, we show that in planar structures the role played by the diamagnetic components becomes negligible, whereas the paramagnetic elements increase their magnetic field channeling efficiency. This is further demonstrated experimentally by tracking the gain of superconductor-ferromagnet concentrators through the superconducting transition. Interestingly, for thicknesses where the diamagnetic petals play an important role for the concentration gain, they also help to reduce the stray field of the concentrator, thus limiting the perturbation of the external field (invisibility). Our findings establish a roadmap and set clear geometrical limits for designing low dimensional magnetic field concentrators.   

%combining paramagnetic materials with diamagnetic components in order to achieve highly anisotropic and spatially modulated magnetic permeability

\end{abstract}

\maketitle

\section{Introduction}

Magnetic field concentrators (MFC) are structures designed to channel and enhance the strength of magnetic fields. They work by guiding the magnetic field lines through a specific path, thereby increasing the magnetic field density in a targeted region. These devices play a vital role in enhancing the sensitivity and efficiency of magnetic sensors and are engineered to manipulate and harness magnetic fields in a way that may serve specific technological needs such as for monitoring power transmission cables \cite{Zhu2019}, magnetoencephalography \cite{kanno2022}, magnetoresistance biosensors \cite{dey2023,chen2022}, and magnetometry based on nitrogen-vacancy quantum probes \cite{Fescenko2020,chen2022,mao2023}. The essential parameters determining the performance of MFC are the material choice and the imposed geometry \cite{Drljaca-2002,Sun-2013,zhang-2018}. Although the vast majority of MFC are based on soft ferromagnetic (FM) alloys, some high field applications have been proposed based on the flux-focusing produced between neighboring superconducting parts \cite{Kiyoshi2009}.

The advent of transformation optics \cite{Pendry} brought about a paradigm shift in the way scientists approach, understand and conceive magnetic field concentrators. This discipline involves a coordinate transformation to control light paths within media. Since Maxwell’s equations are form-invariant to coordinate transformations, only the components of the permittivity tensor and the permeability tensor are affected by the transformation. This approach has served as a bridge between theoretical physics and practical engineering, providing a platform to design electromagnetic devices with unprecedented functionalities and efficiencies \cite{Schurig}. In the limit of very low frequencies, where the electric and magnetic fields become separable in Maxwell’s equations \cite{Wood}, the fundamental ingredient needed to achieve efficient guidance of magnetic field lines and negligible external distortion of a uniform applied field, is a highly anisotropic magnetic permeability tensor $\mu$ \cite{sanchez2011}. 

More precisely, this condition can be fulfilled in an axially symmetric structure by combining a radial ($\mu_{\rho}$) and angular ($\mu_{\theta}$) relative permeability components fulfilling the
relations $\mu_{\rho}\mu_{\theta}=1$ and $\mu_{\rho} \gg \mu_{\theta}$ \cite{sanchez2011}. Recognizing that natural materials do not exist that satisfy these conditions, scientists have proposed several metamaterials, constructed from alternating layers or wedges, to serve as approximations. These materials strategically combine wedges of superconductors to suppress the azimuthal permeability and ferromagnetic wedges to enhance $\mu_{\rho}$ and thus tailor their effective permeabilities to enable the desired electromagnetic behavior. It has been theoretically and experimentally shown that a long cylindrical shell made of such a metamaterial with inner and outer radii $R_i$ and $R_o$, respectively, is able to enhance the magnetic field in the sensing area, by a factor $R_o/R_i$ \cite{navau-2012,Prat-Camps-2014,Navau-2017}. Similar structures combining ferromagnetic and conducting materials have even shown potential for concentrating alternating magnetic fields \cite{Kibret2016}.

The continuous strive for high density and on-chip integrated devices working at room temperature has motivated the investigation of downscaling the metamaterial shells to realize planar meso/micro-metasurfaces made of only magnetic materials \cite{Fourneau2023}. In this context, it has been shown that the concentration power depends on the thickness of the device and achieves an optimal performance when the thickness is about twice the inner radius of the MFC\cite{Fourneau2023}. In addition, the presence of magnetic domains, the irreversible magnetic response, and the saturation of the MFC have shown to have a detrimental impact on the performance of the concentrator by reducing the range of magnetic fields over which the device can operate\cite{Fourneau2023}. To date, it remains unclear whether it is possible to further enhance the gain of 2D magnetic concentrators by inserting highly diamagnetic petals, similar to the successful strategy already demonstrated for 3D devices \cite{Prat-Camps-2014}.

In this work, we address this question by numerically investigating the effect of diamagnetic petals sandwiched between neighboring paramagnetic petals on devices spanning the whole range of possible thicknesses. For MFC with thicknesses larger than the outer radius ($t > R_o$) the presence of diamagnetic petals gives rise to more than 100\% enhancement of the concentration gain. By reducing their thickness, the paramagnetic petals become more efficient by collecting additional magnetic field lines from to the top and bottom surfaces, although the overall gain of the device decreases due to the underperformance of the diamagnetic petals. For thicknesses smaller than the internal radius ($t < R_i$), the diamagnetic petals play a negligible role in the concentration factor and could be simply omitted. We have experimentally confirmed this thin film limit for microscale planar concentrators made of a soft ferromagnet (permalloy) combined with superconducting petals made of a high temperature superconductor (YBa$_2$Cu$_3$O$_7$). We have also demonstrated that adding a magnetic disk with a slit in the core of the concentrator leads to a boost in the concentration effect. In the thickness regime where the diamagnetic petals boost the gain of the structures, they also promote the invisibility of MFC. 

\section{Results and Discussion}

\subsection{Finite Element Modeling}

\begin{figure*} [t!]
\centering
  \includegraphics[width=1\linewidth]{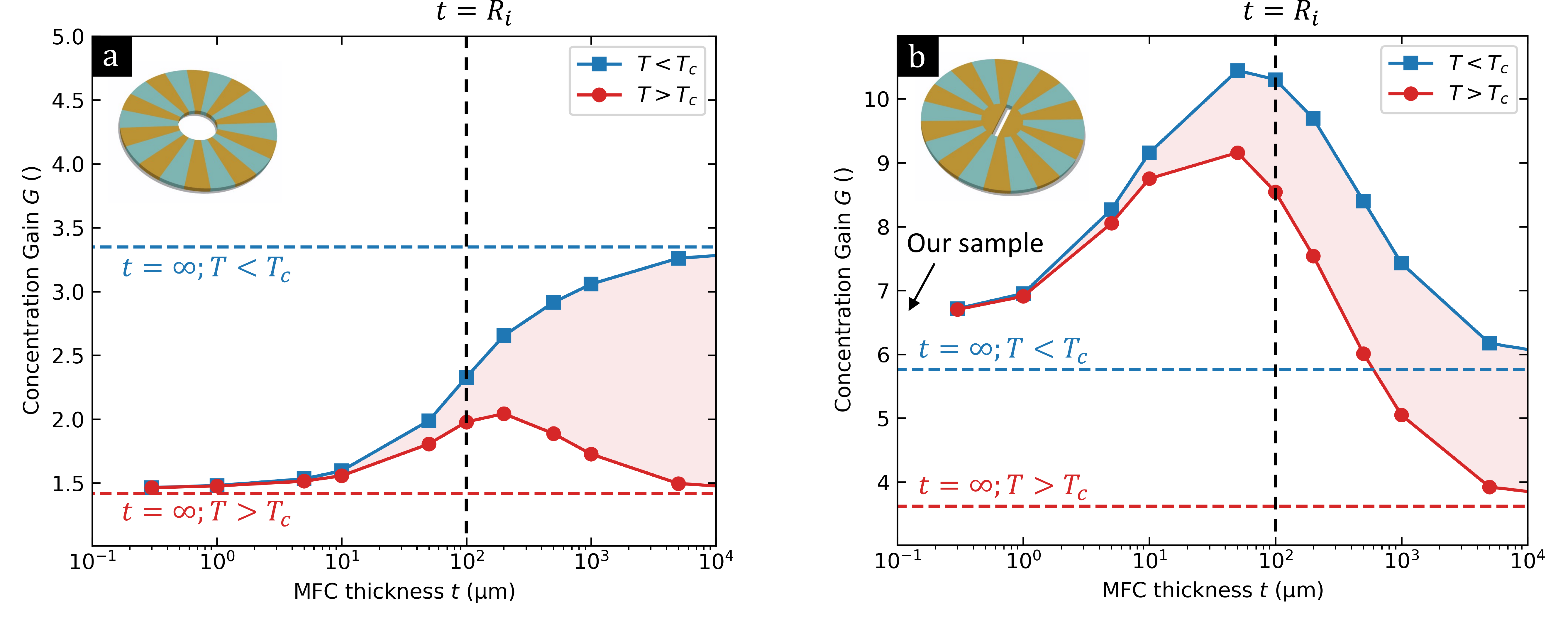}
  \caption{Simulation of the variation in concentration gain with the MFC thickness for concentrators of outer radius $R_o=400~\upmu$m and inner radius $R_i=100~\upmu$m. The simulation is repeated with diamagnetic petals having a relative permeability $\mu_r$ of both 1 and $10^{-5}$ to mimic superconducting petals above and below their critical temperature, $T_c$. Panels (a) and (b) illustrate the differences in results obtained with an MFC, with panel (a) featuring an MFC without a central core and panel (b) depicting results with a ferromagnetic (FM) core.}
  \label{MFC_thickness}
\end{figure*}

The MFC under consideration follow the design proposed in Ref. \onlinecite{Prat-Camps-2014} and consist of a shell with inner radius $R_{i}=100~\upmu$m and outer radius $R_{o}=400~\upmu$m with alternating ideal paramagnetic and diamagnetic petals, as schematically shown in the inset of Fig.\ref{MFC_thickness}(a). In this model, the ideal paramagnetic material has an exceptionally high permeability and no saturation magnetization. This behavior corresponds to the magnetic response of a soft ferromagnetic material, such as permalloy, in a linear regime, specifically at low applied external fields. Conversely, the ideal diamagnetic material is described by a superconductor in the Meissner state, with a critical temperature $T_c$. In this context, we can directly compare the response of the MFC with diamagnetic petals ($T<T_c$) and that of the same concentrator without diamagnetic petals ($T>T_c$). Note that the assumption of the superconductor as a perfect screening components represents an upper bound on the size of any effects. In films with thickness comparable or smaller to the superconducting penetration depth, the field will be able to penetrate from top and bottom. Hence any screening of the in-plane field is going to be substantially less than this in reality.

Simulations are conducted using a finite element method with the system discretized into a tetrahedral mesh grid. The stationary Maxwell's equations in the absence of charge currents are solved using the magnetostatic module of Comsol Multiphysics software, with a relative accuracy set to 10$^{-5}$. The boundary condition dictates that the local magnetic field $\mathbf{B}$ equals the applied field $\mathbf{B_a}$ on each boundary of the simulation box. The dimensions of these faces are sufficiently large to ensure that the error in gain resulting from field line confinement remains below 1\%. The relative permeability values for the ideal paramagnetic and diamagnetic materials are specified as $10^5$ and $10^{-5}$, respectively. These values have been chosen in such a way that the radial and angular magnetic permeabilities, $\mu_r$ and $\mu_\theta$, satisfy the relation $\mu_r \mu_\theta=1$ necessary for avoiding perturbations of the magnetic field around the device \cite{prat-camps-2013}.

Fig.\ref{MFC_thickness}(a) shows the concentration gain $G=\dfrac{\mathbf{B}_0\cdot\mathbf{B}_a}{|B_{a}|^2}$, with $\mathbf{B}_0$ being the field at the inversion symmetry point of the structure (i.e. its center), as a function of thickness. The red (blue) data points correspond to the device without (with) diamagnetic petals. The dotted lines indicate the associated asymptotic limits for an infinite cylindrical sample as calculated in Ref.\onlinecite{prat-camps-2013}. For $t > 10^4~\upmu$m (i.e. $t/R_i > 100$) the response is nearly that of an infinite cylindrical device. For the MFC with only paramagnetic petals, as $t$ decreases, $G$ increases. The reason for this effect is that the additional magnetic field lines entering through the upper and lower surfaces of the paramagnetic petals aligned with the applied field, further contribute to the gain. In striking contrast to this, for the MFC with diamagnetic petals, $G$ decreases as $t$ decreases. This behavior results from the fact that for thin structures ($t/w < 1$, with $w$ the average width of the petal) the expulsion of field lines by the diamagnetic petals towards the top and bottom surfaces is favored over that towards the sides. Note that the performance increase of the paramagnetic petals does not compensate the underperfomance of the diamagnetic petals and an overall decrease of the gain is observed as the thickness decreases. For the MFC with only paramagnetic petals, an optimum gain is observed when $t/R_i \sim 2$, below this ratio the stray field at the tip of the petals rapidly fans out leading to a net decrease of the gain. It is worth noting that for $t < 10~\upmu$m the presence of diamagnetic petals becomes irrelevant for this geometry.

In an attempt to reach higher values of concentration, we investigate the response of a similar device in which we have filled the inner core with a paramagnetic material with a $20~\upmu$m wide slit (see inset of Fig.\ref{MFC_thickness}(b)). The results for an in-plane applied field  perpendicular to the slit are presented in Fig.\ref{MFC_thickness}(b). The thickness dependence of the gain for this MFC follows a similar trend to that of the empty-core MFC but with a more pronounced enhancement at intermediate thicknesses. Although the diamagnetic petals seem to contribute in a larger range of thicknesses, their role eventually becomes negligible for $t < 5~\upmu$m. 

\begin{figure*} [t!]
\centering
  \includegraphics[width=1\linewidth]{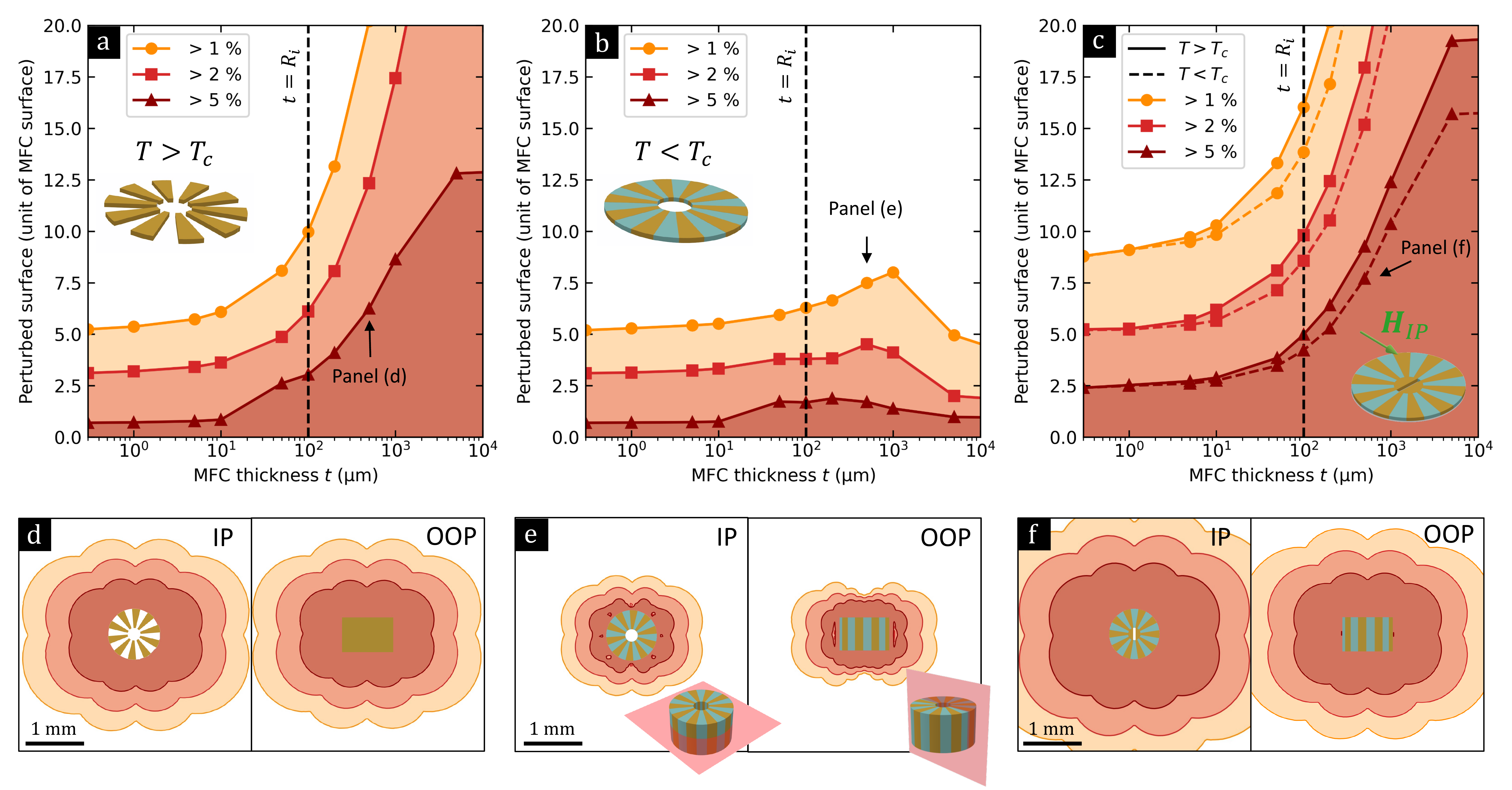}
  \caption{Magnetic field perturbation $|\mathbf{B}-\mathbf{B}_a|/B_a (\%)$ as a function of the MFC thickness. Evolution of the footprint area in the vicinity of the MFC for which the perturbation is larger than 5, 2 and 1 $\%$, (a) for a MFC without FM core above $T_c$, (b) without core below $T_c$ and (c) in the presence of a FM core below and above $T_c$. (d-f) Top and side view of the perturbation areas calculated for 500-$\upmu$m-thick devices.}
  \label{MFC_invisibility}
\end{figure*}

Another intriguing property of flower-like magnetic metamaterial concentrators, which has been theoretically predicted and experimentally demonstrated for macroscopic structures, is the confinement of the stray field in the close vicinity of the concentrator, rendering it magnetically invisible. In other words, for perfectly invisible concentrators, the magnetic field around the MFC exactly matches the applied field \cite{prat-camps-2013}. In Fig. \ref{MFC_invisibility}, we simulate the impact of the MFC on the magnetic field landscape, taking into account the thickness of an 8-petal-concentrator. To assess the magnetic invisibility, we calculate the projected area around the MFC within which the local magnetic field deviates by no more than a few percent of the applied field, i.e., $\epsilon=\dfrac{|\mathbf{B}-\mathbf{B}_a|}{B_a} < 1, 2, 5 \%$.

In the regime above the critical temperature of the device (Fig. \ref{MFC_invisibility}(a)), the area affected by the MFC increases linearly with the sample thickness (the abscissa is expressed on a logarithmic scale). For instance, the 1\% perturbation region covers a surface that grows from 5 times the concentrator's surface when $t=1~\upmu$m to more than 20 times for 0.5-mm-thick samples. For the sake of clarity, the perturbed area for the specific case of $t=500~\upmu$m is displayed in panel (d) for both an in-plane (IP) and an out-of-plane (OOP) cross-sectional view. Both cuts are symmetry planes of the device as shown in the inset of panel (e).

When $T<T_c$ (Fig. \ref{MFC_invisibility}(b)), the presence of the diamagnetic petals drastically reduces the extent of the stray field which remains confined within a radial distance of $2R_i$ from the edge of the MFC. Note that a shallow maximum of the perturbed fields outside the MFC develops when $t$ is about $2R_o$. The improved invisibility of the MFC with diamagnetic petals becomes more apparent by comparing panels (d) and (e), corresponding to the absence and presence of diamagnetic petals, respectively. However, it should be noted that, similar to the gain improvement shown in Fig. \ref{MFC_thickness}, negligible difference is observed when the thickness of the device falls below $t=10~\upmu$m. Finally, for the concentrator with a ferromagnetic core (panels (c) and (f)), the difference between having (circles connected by dashed lines) or not having (squares connected by lines) diamagnetic petals, remains negligible even for the thicker samples.

These results suggest that diamagnetic petals have little influence on the gain of thin MFC but may offer the beneficial effect of boosting the concentration gain and improving their invisibility for thick devices, thus reducing their influence on neighboring electromagnetic components of the chip. In the next section we address these aspects from an experimental point of view.

\subsection{Experimental results}

The theoretical predictions for an infinitely long cylindrical MFC based on a model of combined ideal paramagnetic and diamagnetic components were experimentally validated in Ref. \onlinecite{Prat-Camps2014}. Here we will explore the opposite extreme and gauge the impact of implementing diamagnetic elements in planar 2D on-chip MFC. 
To this end we have fabricated sub-millimeter MFC alternating permalloy (Py) petals with superconducting petals made of a high-temperature superconductor YBa$_2$Cu$_3$O$_{7-\delta}$ (YBCO). The choice of YBCO is well justified by the fact that a strong diamagnetic response is expected due to its high lower critical field for IP magnetic fields ($\mu_0 H_{c1} \approx 23$ mT)\cite{Moshchalkov} and at the same time its high critical temperature ($T_c\approx87$ K) offers a wide and easily accessible temperature range for exploring different diamagnetic regimes. 

\begin{figure*}[t!]
  \includegraphics[width=\linewidth]{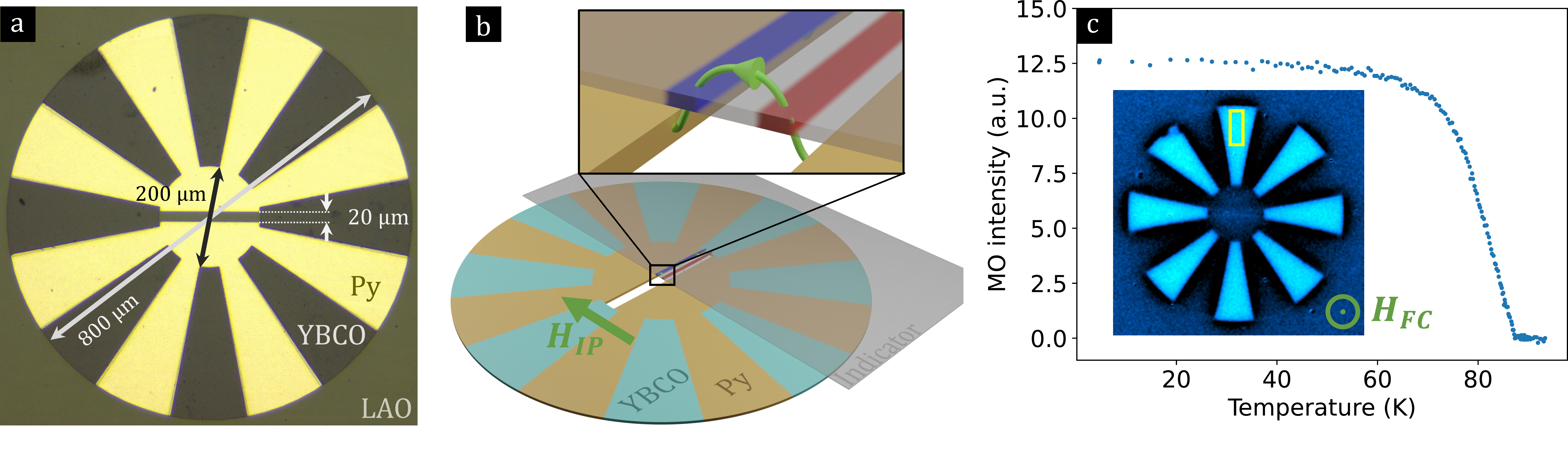}
  \caption{(a) Optical image of the investigated sample. The outer radius is four times larger that the inner one. (b) Schematic view of the experimental setup used to extract the polarization at the gap via magneto-optical imaging (MOI). A Faraday indicator is placed on top of the sample mounted in a cryostat and the in-plane magnetic field is swept while capturing images of the magnetic landscape for various temperatures. The indicator senses the out-of-plane component of the magnetic field coming from the stray field above the gap, as depicted in the close-up view. (c) Evolution of the MOI signal arising from trapped flux in the YBCO after field cooling in an out-of-plane field as a function of the temperature. The mean value in the yellow rectangle from which the slightly temperature-dependent background was subtracted gradually drops to zero at $T_c = 87$ K.}
  \label{YBCO}
\end{figure*}

\begin{figure*}[t!]
    \centering
    \includegraphics[width=\linewidth]{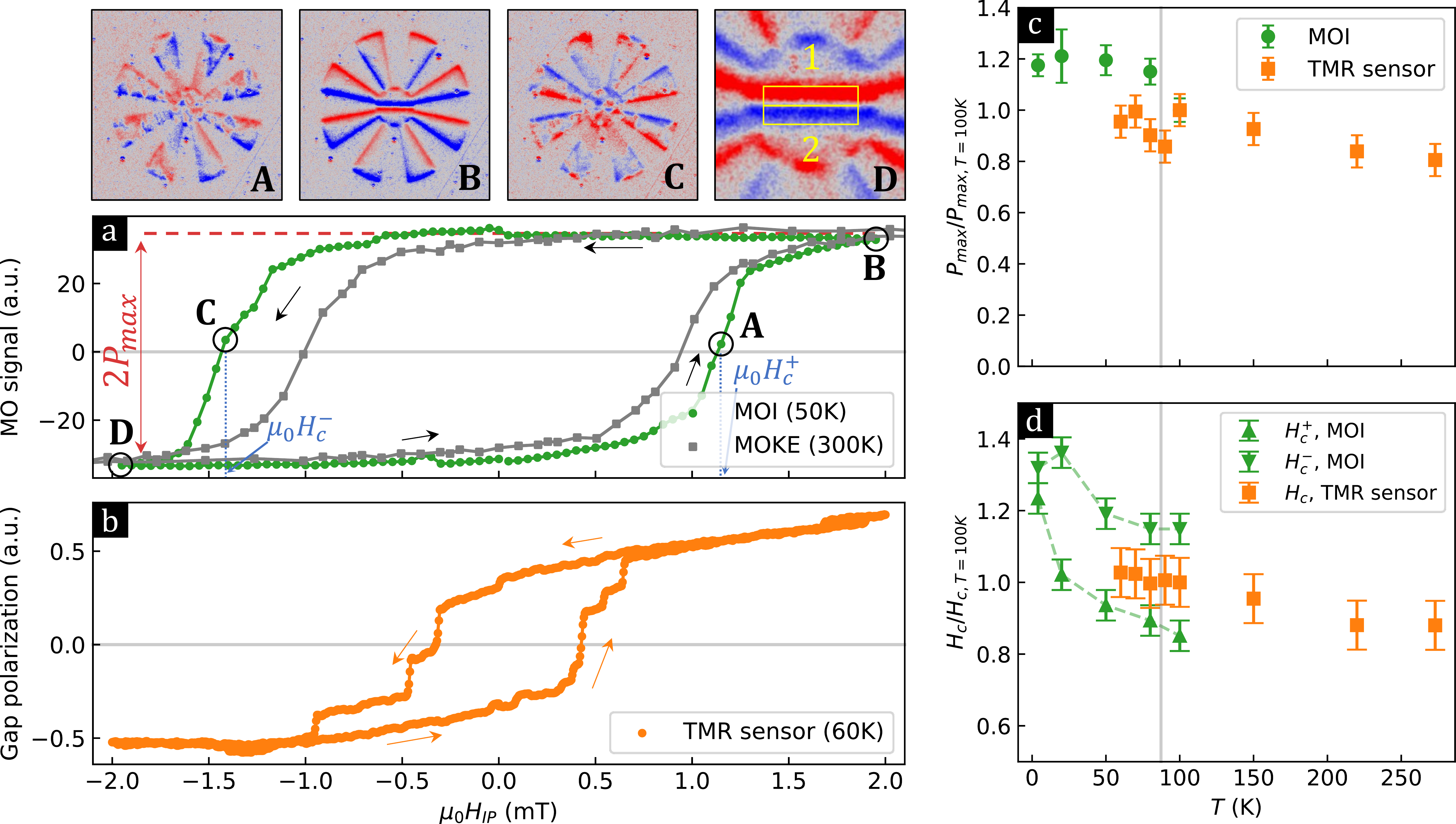}
    \caption{(a) Representative hysteresis curve of the OOP component of the stray field at the gap at $T=50$ K, extracted from MOI sweeps (green circles). The gap polarization $P$ is defined as the difference in the mean pixel value between region 2 and region 1, both shown in image D. MOI images at different magnetic fields close to saturation or polarization reversal are shown in panel A-D. The hysteresis loop starts at 0 mT in the negative remnant configuration. Gray square symbols show the hysteresis loop obtained on the same sample by tracking the longitudinal magnetic Kerr signal at room temperature and sweeping the magnetic field between $-8$ mT and $+8$ mT. (b) Magnetic hysteresis loop at $T=60$ K measured with a TMR sensor at the edge of the slot in an 8-petal Py/YBCO concentrator (see text). (c) Relative evolution of the maximum polarization with temperature for both MOI and scanning TMR sensor techniques. The maximum polarization is defined as half the difference between polarization at maximum and minimum fields. (d) Temperature dependence of the normalized in-plane coercive field at which the magnetic polarization reverses. For MOI, both field directions are differentiated. Dashed lines are guides to the eye and the vertical gray line is at $T_c\approx87$ K.}
    \label{fig:fig4}
\end{figure*}

%The sample layout consists of a $5 \times 5~$mm$^2$ LaAlO$_3$ (LAO) substrate with .  
Five different devices having an external radius $R_{o}=400~\upmu$m and petals touching a central Py disk of $R_{i}=100~\upmu$m were fabricated by growing a 100 nm-thick YBCO film on a $5\times 5$ mm$^2$ (001)-LaAlO$_3$ substrate by pulsed laser deposition at 800\textdegree C under an O$_2$ partial pressure of $0.3$ mbar. The YBCO petals were then defined by standard photolithography and wet etching. Subsequently, the Py parts were fabricated by DC magnetron sputtering and lift-off. An optical image depicting one of the devices is shown in Figure \ref{YBCO}(a). The thickness of the YBCO and Py layers is 100 nm thus avoiding the formation of stripe domains in the Py film. All devices have a slit-shaped gap of width $g$ in the center, allowing us to pick up the OOP components of the stray field (as schematically represented in Figure \ref{YBCO}(b)) which is proportional to the magnetic field concentration gain. A device with no superconducting components and $g=5~\upmu$m is used as a reference. Four other devices, two of them with 16 petals (half superconducting, half ferromagnetic) and gaps of 5 and $20~\upmu$m, and another two with 4 petals (half superconducting, half ferromagnetic) and gaps of 5 and $20~\upmu$m, were investigated and all exhibit similar behavior.

The MFC were investigated through quantitative magneto-optical imaging (MOI) based on the Faraday rotation of an indicator film placed on top of the device. Details of the technique and the setup can be found in Ref. \onlinecite{Shaw2018}. In order to investigate the possible reduction of critical temperature of the superconducting components after microstructuring, we cooled down the sample to the base temperature of our cryostat ($\sim$ 4 K) while applying an OOP magnetic field $\mu_0 H_{\text{FC}}=4$ mT, subsequently the field is turned off and the trapped flux is monitored as a function of temperature. Figure \ref{YBCO}(c) shows the background subtracted intensity recorded in the rectangular box indicated in the inset, as a function of temperature. A superconducting critical temperature $T_c \approx 87$ K is obtained for all superconducting petals. The obtained sequence of images as a function of temperature shows that the magnetic response of the superconducting petals is uniform throughout the entire device (see animation in the Supplemental Material). 

The investigation of the MFC exposed to an IP field requires a very specific image post-processing protocol. Indeed, in view of the fact that the FM elements exhibit a remnant magnetization, standard image processing involving background subtraction would require one to heat up beyond the Curie temperature of Py ($\sim$ 350$^\circ$C). Here we introduce an alternative method consisting of averaging the magnetic images corresponding to the two opposite polarities of the maximum IP applied magnetic field, $\pm \mu_0 H_{\text{max}}=1.95$ mT. By doing so, the stray field emanating from the FM cancels out, and only the illumination background remains. The result can then be subtracted from the acquired images to reveal the stray field of the sample. 

The OOP component of the stray field in the gap region can then be recorded as a function of the applied IP field and temperature as shown in Figure \ref{fig:fig4}(a). By computing the mean intensity value in the regions 1 and 2 indicated by the yellow frame in picture D of Figure \ref{fig:fig4}, the hysteresis loop shown in the panel (a) is obtained as the IP field is cycled. Note that the magnetically active indicator has a nearly linear response to OOP fields up to 125 mT (well above the maximum stray field $\sim 10$ mT) ensuring good proportionality between light intensity and the OOP component of the stray field. 
As temperature decreases, both the saturation polarization (green circles in panel (c)) and the coercive field (green triangles in panel (d)) increase, in agreement with previous reports on Py films \cite{Luo2015,Zhao2016}. Note, however, that no particular feature or behavior is observed when the YBCO switches from a non-magnetic ($T=100$ K $> T_c$) to strongly diamagnetic ($T=80$ K $< T_c$) state. More precisely, the superconducting transition in the YBCO elements does not seem to induce an enhancement of the magnetic field concentration efficiency. Further measurements comparing a MFC with and without diamagnetic petals corroborate this result. These findings are consistent with the results of finite-element simulations presented in Fig.\ref{MFC_thickness}(b), showing that for the maximum IP applied field (2 mT), the difference between having and not having perfectly diamagnetic petals is about $1~\upmu$T, well below our experimental resolution. 

Note that the hysteresis loop in Figure \ref{fig:fig4}(a) is not centered at $\mu_0 H=0$ but it is biased to negative fields. This behavior arises due to the fact that the maximum applied IP field is not enough to completely saturate the magnetic moment of the MFC. The noisy response in the vicinity of $\mu_0 H=0$ results from the inevitable formation of magnetic domains in the indicator used to capture the MOI images. The MOI images corresponding to points A-D in panel (a) show the OOP component of the stray field and reveal the proliferation of magnetic domains at the coercive field. 

Further information concerning the magnetic response of the MFC has been obtained through two complementary techniques. Firstly, magneto-optical Kerr effect (MOKE) microscopy allowed for tracking the evolution of the IP magnetic domains in the FM petals at room temperature. We use a commercial MOKE microscope from Evico magnetics in the longitudinal Kerr configuration using 20$\times$ and 10$\times$ magnification objectives \cite{moke}. MOKE images presented in this work are obtained employing background subtraction. A snapshot of the magnetic domains near the coercive field of the structure is presented in Annex I and an animation of the MOKE images as a function of field can be found in the Supplemental Material. By computing the average intensity signal of the entire device, it is possible to obtain a hysteresis loop as shown by the gray square symbols in Figure \ref{fig:fig4}(a). The lower coercivity of this loop with respect to that obtained by MOI results from the temperature dependence of the $\mu_0 H_c$, as shown in panel (d). Secondly, we compare to low-temperature scanning tunnel magnetoresistance (TMR) microscopy, which is a non-invasive technique permitting to either scan an area to reveal the OOP magnetic landscape or acquire the local response by parking the sensor in a particular spot. A local hysteresis loop is presented in Figure \ref{fig:fig4}(b) and scan areas are available in the Annex I.  Note that the magnetization loop obtained by this method exhibits a substantially smaller coercivity. This difference might be attributed to the fact that the coercive field is position dependent, as revealed by the MOKE measurements presented in Annex I. The staircase shape of the TMR loop shown in Figure \ref{fig:fig4}(a) can be naturally explained by the process of depinning and reversal of adjacent magnetic domains. Annex II summarizes the hysteresis loops obtained by TMR measurements for several devices and as a function of temperature.

\section{Conclusions}
We have numerically investigated the performance of magnetic field concentrators based on cylindrical metamaterial shell structures, as a function of their thickness for fixed inner and outer radii. Two figures of merit are introduced to evaluate the performance of the devices, the gain and the invisibility. For MFC without diamagnetic petals, the gain is optimized when the thickness is comparable to the inner radius of the shell. This maximum arises from the competition between the increased dispersion of the stray field emanating from the tips of the petals as the thickness decreases and the enhancement of the collected field lines through the upper and lower surfaces as the thickness decreases. We show that the effect of the diamagnetic petals on the gain rapidly becomes negligible as the thickness decreases, but they play an important role in limiting the perturbation of the field outside the structure. We have experimentally investigated the thin-film limit in devices composed of alternating superconducting and ferromagnetic petals, and confirm the predictions of the numerical model. The obtained results provide clear guidelines for designing low dimensional magnetic field concentrators.     

% Acknowledgements
%\medskip
\begin{acknowledgments}
% \section{Acknowledgements}\par
This work was supported by the Fonds de la Recherche Scientifique - FNRS under the programs PDR T.0204.21 and CDR J.0176.22, EraNet-CHISTERA R.8003.21, the Spanish Ministry of Science and Innovation MCIN/ AEI /10.13039/501100011033/ through CHIST-ERA PCI2021-122028-2A and PCI2021-122083-2A cofinanced by the European Union Next Generation EU/PRTR, HTSUPERFUN PID2021-124680OB-I00 cofinanced by ERDF a way of making Europe and FIP-2020 METAMAG and PID2019-104670GB-I00 of the Agencia Estatal de Investigación/Fondo Europeo de Desarrollo Regional (UE), and by COST (European Cooperation in Science and Technology) [www.cost.eu] through COST Action SUPERQUMAP (CA 21144). Access to the CEITEC Nano Research Infrastructure was supported by the Ministry of Education, Youth, and Sports (MEYS) of the Czech Republic under the project Czech NanoLab (LM2023051). J.A.A. and V.U. acknowledge the support from the TACR EraNet CHIST-ERA project MetaMagIC TH77010001. S. J. B. was supported by the Engineering and Physical Sciences Research Council (EPSRC) in the United Kingdom under Grant No. EP/W022680/1. N. L. acknowledges support from FRS-FNRS (Research Fellowships FRIA). The work of E. Fourneau has been financially supported by the FWO and F.R.S.-FNRS under the Excellence of Science (EOS) project O.0028.22. 

N. Lejeune and E. Fourneau contributed equally to this work.

\end{acknowledgments}

\section*{Conflict of Interest}
The authors have no conflicts to disclose.
\section*{Data Availability Statement}

The data that support the findings of this study are available from the corresponding author upon reasonable request.
\appendix

% Change the figure notation and number
\setcounter{figure}{0} 
\renewcommand\thefigure{S\arabic{figure}}  

\section{Magneto-optical Kerr microscopy analysis}

\begin{figure}[h!]
    \centering
    \includegraphics[width=\linewidth]{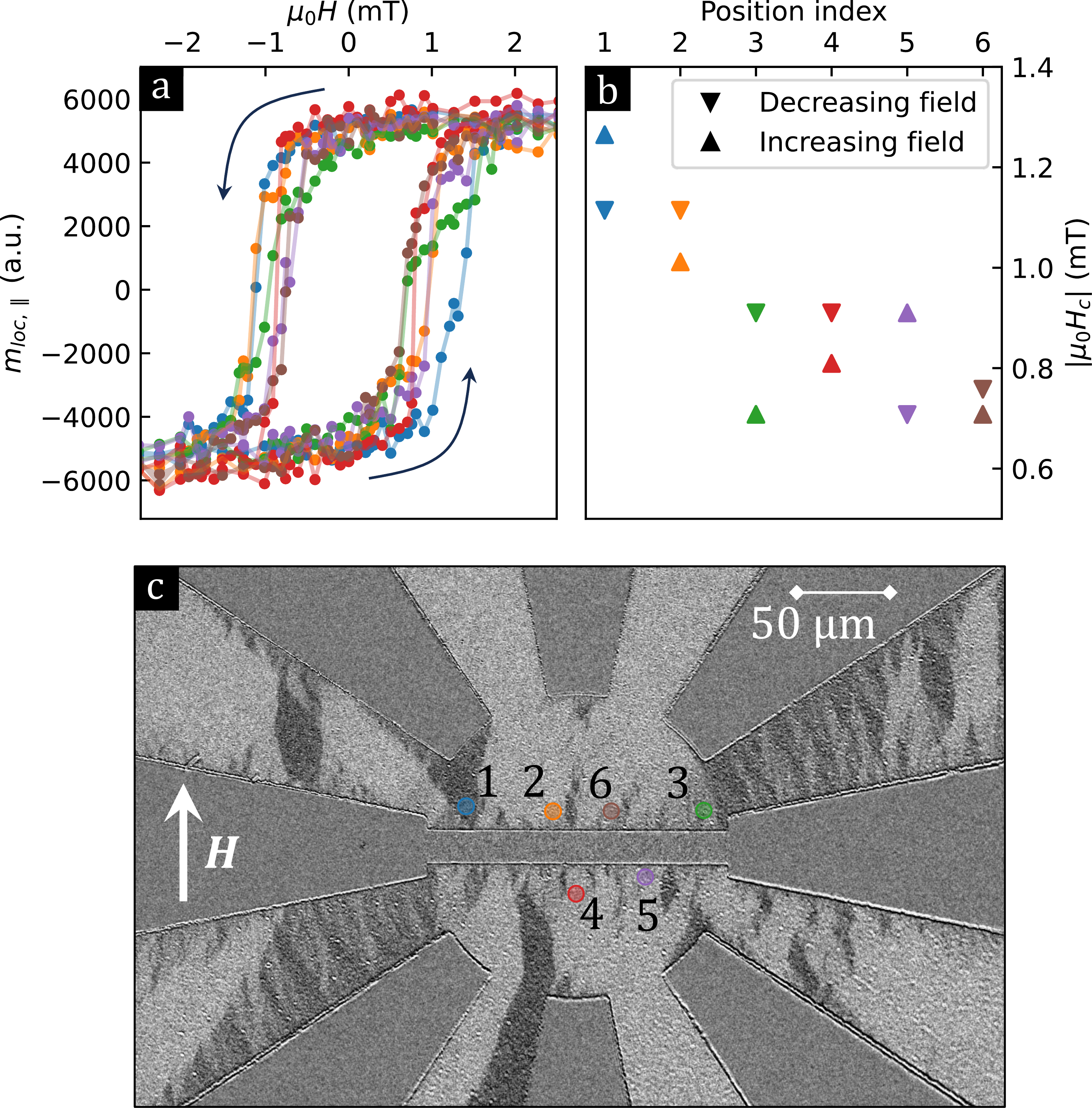}
    \caption{(a) Hysteresis loops obtained by averaging the MOKE signal at the spots indicated in panel (c). The color code of the loops relates to the colors of the regions indicated in (c). The applied magnetic field is swept in the vertical direction, i.e. perpendicular to the gap. Measurements are acquired at room temperature. (b) Extracted coercive field as a function of the location of the spot indicated in (c). Panel (c) shows the longitudinal MOKE signal (Kerr axis vertical) at $\mu_0 H = 1.1$ mT. The observed color contrast corresponds to variations of the vertical magnetization $m_y$ thus revealing the magnetic domains.}
    \label{fig:S1}
\end{figure}

MOKE microscopy images at room temperature were acquired while cycling magnetic field between $-8$ mT and $+8$ mT, both in longitudinal and transversal mode. The particular benefit of this technique is that it allows one to locally map the IP magnetization of the MFC (in contrast to MOI and TMR which pick up the OOP component of the stray field). Thus, magnetic domains become visible and their evolution as a function of the applied IP field can be tracked. From the average intensity of these images over a chosen area it is also possible to reconstruct magnetic hysteresis loops. Figure \ref{fig:S1}(a) shows hysteresis loops obtained at different regions of the MFC labeled 1-6 in panel (c) for an IP magnetic field perpendicular to the gap. Panel (c) shows the longitudinal MOKE signal (Kerr sensitivity with respect to the vertical magnetization component) at $\mu_0 H = 1.1$ mT. The observed color contrast corresponds to variations of the vertical component of the local magnetic moment $m_y$ thus revealing the magnetic domains. Note that in panel (a), saturation and coercivity are position dependent properties. This is more apparent in panel (b) showing the coercive field as a function of the labeled location indicated in panel (c). Note that a variation of $\mu_0 H_c$ as large as a factor of 2 can be obtained on the same device. An animation of the evolution of the IP magnetic domain distribution during the magnetization reversal process is shown in the Supplemental Material.  

\section{Scanning TMR imaging}

\subsection{Experimental details}

Hysteretic magnetization loops were acquired with a commercial TMR sensor fabricated by Micro Magnetics Inc. A $2 \times 4~\upmu$m$^2$ elliptical junction has been patterned $\sim 5~\upmu$m from the lapped corner of a Si substrate and the sensor exhibits approximately linear sensitivity to OOP stray magnetic fields from the sample ($\Delta V=1.4654$ mV/mT with a $10~\upmu$A, 1.032 kHz drive current). To measure the magnetization, a 3-axis piezoelectric stage was used to position the TMR sensor just above the center of the $20~\upmu$m slot at a point near the edge where the OOP component of the field was maximum. The IP magnetic field generated by a Helmholtz coil pair was then swept in a loop with extrema $\mu_0 H_{\text{IP}} = \pm2$ mT while the sensor voltage was recorded. During measurements the sample was mounted on the cold head of a low vibration commercial cryocooler inside an evacuated cryostat, allowing the temperature to be stabilized in the range 60-300 K.

\subsection{Hysteretic response}

\begin{figure*}
    \centering
    \includegraphics[width=\linewidth]{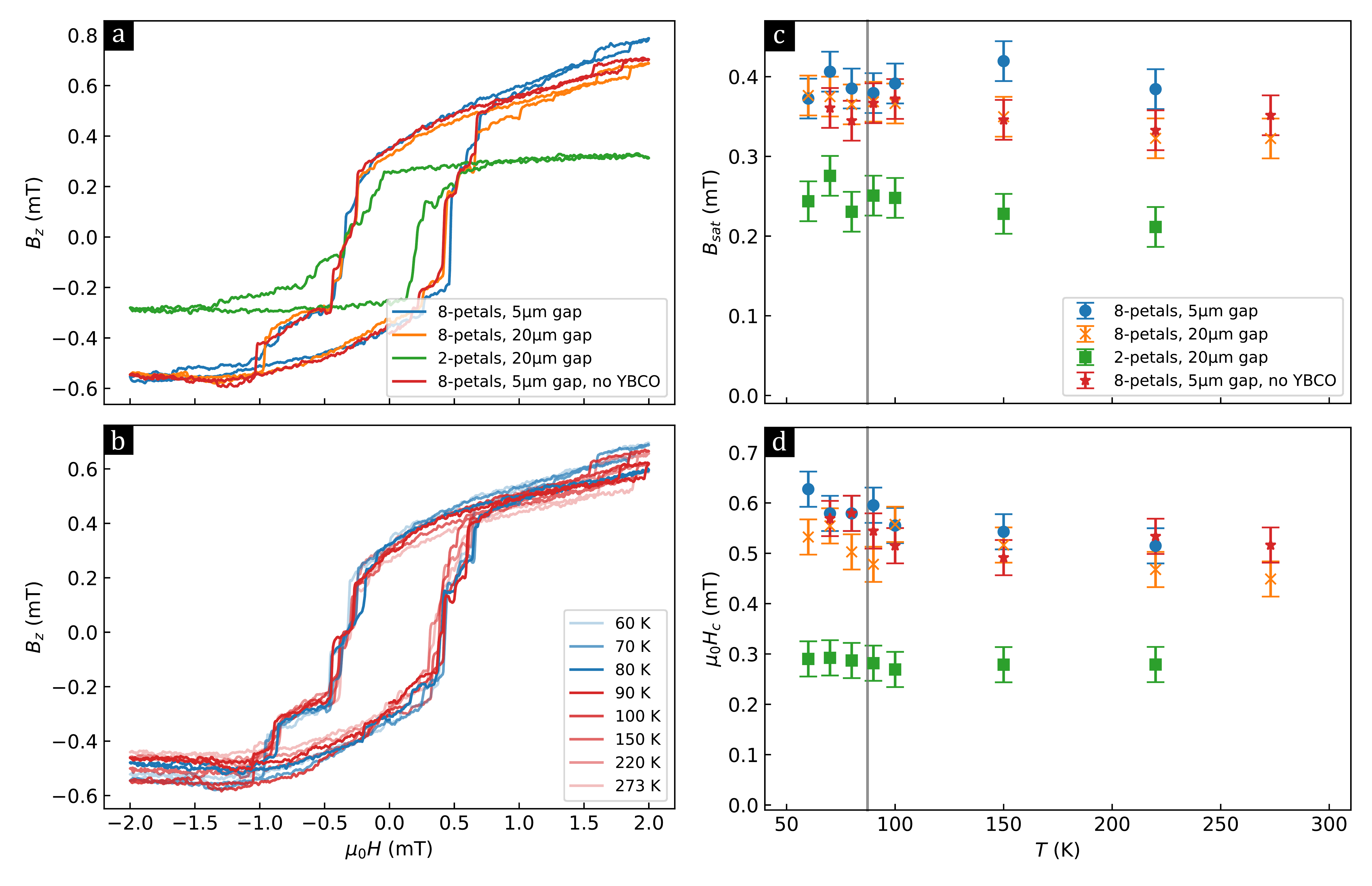}
    \caption{(a) TMR sensor magnetic hysteresis loops for 4 different devices at $T=70$ K. (b) Hysteresis loops for various temperatures for an 8-petal Py/YBCO concentrator structure with a 20-$\upmu$m-wide slot. No detectable change is observable between data obtained below (blue) and above (red) the superconducting transition $T_c = 87$ K. (c) Saturation field and (d) coercive field as a function of temperature extracted from the hysteresis loops for various devices. The vertical gray line indicates the superconducting transition.}
    \label{fig:supplBath}
\end{figure*}

Figure \ref{fig:supplBath}(a) presents hysteresis loops for different devices recorded using the TMR sensor at $T=70$ K for IP magnetic fields between $-2$ mT and $+2$ mT. Note that the 2-petal device exhibits a lower coercive field than the 8-petal MFCs. No significant differences among the various 8-petal devices are observed, neither for those including YBCO petals nor for different gap sizes. Fig. \ref{fig:supplBath}(b) presents the hysteresis loops of the 8-petal devices for several temperatures. No detectable change is observed at the superconducting transition. The extracted saturation fields (i.e. the field $B_z$ measured at $-2$ mT) and coercive fields for each temperature and device are compiled in Fig. \ref{fig:supplBath}(c) and (d), respectively. Panel (b) shows the measured OOP saturation field, $B_{\text{sat}}$, at the edge of the gap as a function of sample temperature. This measurement should be a good proxy for the temperature-dependent saturation magnetization, $M_{\text{sat}}$, near the center of the Py concentrator. Fig. \ref{fig:supplBath}(d) shows the temperature dependence of the measured IP coercive field, $\mu_0 H_c$. As temperature decreases, we observe a weak increase in both the saturation field (panel (c)) and the coercive field (panel (d)) consistent with previous reports on Py films\cite{Luo2015, Zhao2016}. Note, however, that there is no indication of a change in behavior of either parameter when the YBCO switches from being a non-magnetic metal at e.g., $T=90$ K ($T> T_c= 87$ K) to a strongly diamagnetic superconductor at e.g., $60$ K ($T<T_{c}$). In particular, the superconducting transition in the YBCO elements does not seem to induce an enhancement of the magnetic field concentration efficiency.

\subsection{Mapping of the out-of-plane component of the stray field}

\begin{figure}
    \centering
    \includegraphics[width=\linewidth]{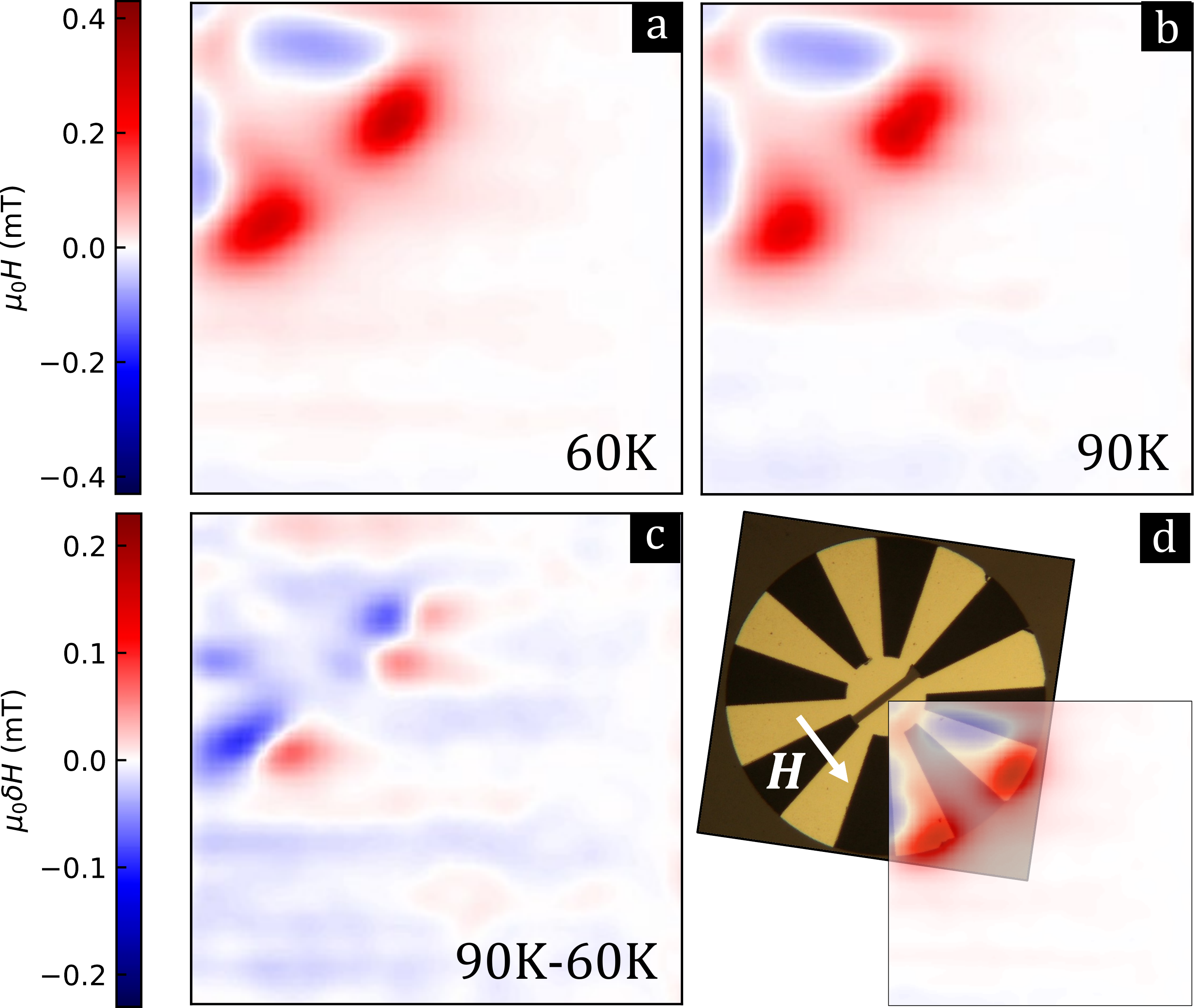}
    \caption{Scanning TMR maps of the out-of-plane component of the magnetic field of an 8-petal Py/YBCO concentrator structure under an in-plane applied field of 2 mT applied perpendicular to the $20~\upmu$m-gap at (a) 60K $<T_c$ and (b) 90K $>T_c$. (c) Pixel-to-pixel numerical difference between the magnetic maps above and below $T_c$. (d) Overlap of the magnetic landscape on top of an optical image of the concentrator. The scanned area corresponds to a region of $750\times750~\upmu$m$^2$.}
    \label{fig:S3}
\end{figure}

Fig. \ref{fig:S3} shows scanning TMR images of one quarter of the concentrator sample studied in Fig. 4. These images were captured using the 3-axis piezoelectric stage to raster scan the sensor across the surface of the sample to generate a $50 \times 50$ pixel map of the OOP field, $B_z$, at temperatures well below (60 K) and above (90 K) the superconducting $T_c$ of the YBCO film. In each case the sample was cooled to the target temperature and an IP of $\mu_0H = 2$ mT applied perpendicular to the $20~\upmu$m slot in the center of the concentrator prior to imaging. The TMR sensor was initially positioned about $25~\upmu$m above the sample surface to prevent any danger of making contact with it during these rather large area scans. Fig. \ref{fig:S3}(a,b) shows maps of the OOP magnetic field, $B_z$, above one quarter of an 8-petal Py/YBCO ‘flower’ structure at (a) 60 K and 90 K (b), well below and above the YBCO superconducting $T_c$, respectively. The measurement geometry shown in panel (d) indicates the position of the scanning area with respect to the MFC. Panel (c) shows the differential image (pixel-to-pixel numerical subtraction) between the magnetic maps above and below $T_c$. We found that when the YBCO petals are in the superconducting state, a slight narrowing and increase in amplitude of the main intensity peaks at the Py poles is observed, along with a weak suppression of the measured field at long distances. Therefore, the YBCO petals act to make the concentrator slightly more invisible at long distances outside its footprint.

\section{Supplemental Material}

\subsection{Animation: MOI out-of-plane trapped magnetic field}

The animation Tc-MOI.mp4 shows the MOI intensity map as a function of temperature of the trapped magnetic flux. The initial state was obtained by cooling down the sample to the base temperature of our cryostat ($\sim$ 4 K) while applying an OOP magnetic field $\mu_0 H_{FC}=4$ mT, subsequently the field is turned off and the magnetic field landscape revealing the field-cooled trapped flux is monitored as a function of temperature.

\subsection{Animation: Field dependent magnetic domain distribution obtained from MOKE microscopy}

The animation MOKE.mp4 shows the MOKE intensity maps for transversal (left) and longitudinal (right) configurations as a function of the IP magnetic field applied vertically. Images were obtained with a 20x objective, the field of view corresponds to $552 \times 345~\upmu$m$^2$. For each magnetic field, 8 images were averaged.
% Slide 13 from Joni

\medskip

% \section{References}\par
% \bibliographystyle{MSP}
% \bibliographystyle{apsrev4-2}

\section*{References}
\bibliography{main}
\end{document}